# Growth of Two-dimensional Compound Materials: Controllability, Material Quality, and Growth Mechanism

Lei Tang[1+], Junyang Tan[1+], Huiyu Nong[1], Bilu Liu[1*], Hui-Ming Cheng[1,2*]

[1]Shenzhen Geim Graphene Center, Tsinghua-Berkeley Shenzhen Institute and Tsinghua Shenzhen International Graduate School, Tsinghua University, Shenzhen, Guangdong, 518055, P. R. China

[2]Shenyang National Laboratory for Materials Sciences, Institute of Metal Research, Chinese Academy of Sciences, Shenyang, Liaoning, 110016, P. R. China

[+]These two authors contributed equally.

Correspondence should be addressed to bilu.liu@sz.tsinghua.edu.cn (BL)

hmcheng@sz.tsinghua.edu.cn (HMC)

CONSPECTUS: Two-dimensional (2D) compound materials are promising materials for use in electronics, optoelectronics, flexible devices, *etc*. because they are ultrathin and cover a wide range of properties. Among all methods to prepare 2D materials, chemical vapor deposition (CVD) is promising because it produces materials with a high quality and reasonable cost. So far, much efforts have been made to produce 2D compound materials with large domain size, controllable number of layers, fast-growth rate, and high quality features, *etc*. However, due to the complicated growth mechanism like sublimation and diffusion processes of multiple precursors, maintaining the controllability, repeatability, and high quality of CVD grown 2D binary and ternary materials is still a big challenge, which prevents their widespread use.

Here, taking 2D transition metal dichalcogenides (TMDCs) as examples, we review current progress and highlight some promising growth strategies for the growth of 2D compound materials. The key technology issues which affect the CVD process, including non-metal precursor, metal precursor, substrate engineering, temperature, and gas flow, are discussed. Also, methods in improving the quality of CVD-grown 2D materials and current understanding on their growth mechanism are highlighted. Finally, challenges and opportunities in this field are proposed. We believe this review will guide the future design of controllable CVD systems for the growth of 2D compound materials with good controllability and high quality, laying the foundations for their potential applications.

1. **INTRODUCTION**

Since 2004, when Geim et al. discovered monolayer graphene using the Scotch tape exfoliation technique,[1] there has been a new era of two-dimensional (2D) materials including metals (*e.g.*, graphene, MXenes, and metallic TMDCs), semiconductors (*e.g.*, black phosphorus, graphitic carbon nitride, metal oxide, semiconducting TMDCs, $Bi_2O_2Se$, and $MoSi_2N_4$), and insulators (*e.g.*, boron nitride (BN), silicates, and mica).

Recently, 2D compound semiconductors have ignited considerable interests due to their wide material choices and possible applications in many areas like as channel materials in 2D integrated electronics. Compared with the traditional silicon and III-V semiconductors (*e.g.*, GaAs, GaN), 2D compound materials have three important advantages. First, the thickness can reach the atomic level and the structure remain stable. Second, they have atomic flatness and no dangling bonds on the surfaces, avoiding the influence of carrier transport due to the trapped states. Third, the ultra-thin thickness makes 2D materials compatible with flexible substrates, facilitating the production of flexible and wearable devices. These advantages make 2D compound materials promising candidates for use in electronics,[2] optoelectronics,[3] flexible devices,[4] *etc.*

To meet the increasing requirements for these applications, materials preparation is the first prerequisite. Up to now, 2D compound materials have been prepared by top-down and bottom-up methods. For top-down method, although the micromechanical exfoliation can be used for a wide range of 2D materials, it is highly dependent on the researcher's experience and the yield is low. Top-down exfoliation is suitable for mass-production of 2D materials, while with small domain sizes.[5] By comparison, bottom-up synthesis by chemical vapor deposition (CVD) has shown its potential for preparing large-area 2D materials with electronic-grade quality and reasonable cost. However, for the growth of 2D compound materials, due to the use of multiple precursors, the complicated vapor-phase growth process has many issues, such as the non-uniform distribution of the as-grown domains, lots of defects and vacancies existed in the samples, as well as its unclear growth mechanism. All these factors should be carefully understood for controlling the growth process of 2D compound materials. Recently, our group has focused on the controlled growth of 2D materials and developed several strategies for this purpose by using different types of precursors and designing CVD reactors to obtain a range of 2D compound materials with good uniformity, high controllability, and high quality.

In this Accounts, taking the growth of TMDCs as an example, we first highlight the important strategies for the controlled growth of 2D TMDCs based on recent progress. Several key factors in the CVD process, including the choice of non-metal and metal precursors, substrate engineering, temperature, and gas flow, are discussed in detail. In addition, the growth mechanism which are related to the nucleation and fast-growth kinetics of TMDCs are addressed. Finally, we put forward challenges and opportunities in this field and potential research directions for the controlled growth of 2D compound materials.

## 2. GROWTH OF 2D COMPOUND MATERIALS

To date, there have been many advances in the preparation of 2D compound materials by CVD. The CVD growth process is highly related to several crucial parameters, including (i) the types of non-metal and metal precursors, (ii) substrate engineering, (iii) temperature, and (iv) gas flow (Figure 1). In addition, the nucleation and growth mechanism of 2D compound materials will be discussed. An in-depth understanding of these parameters and the growth mechanism is important to achieve the controlled synthesis of 2D compound materials, including binary, ternary to even more complicated materials.

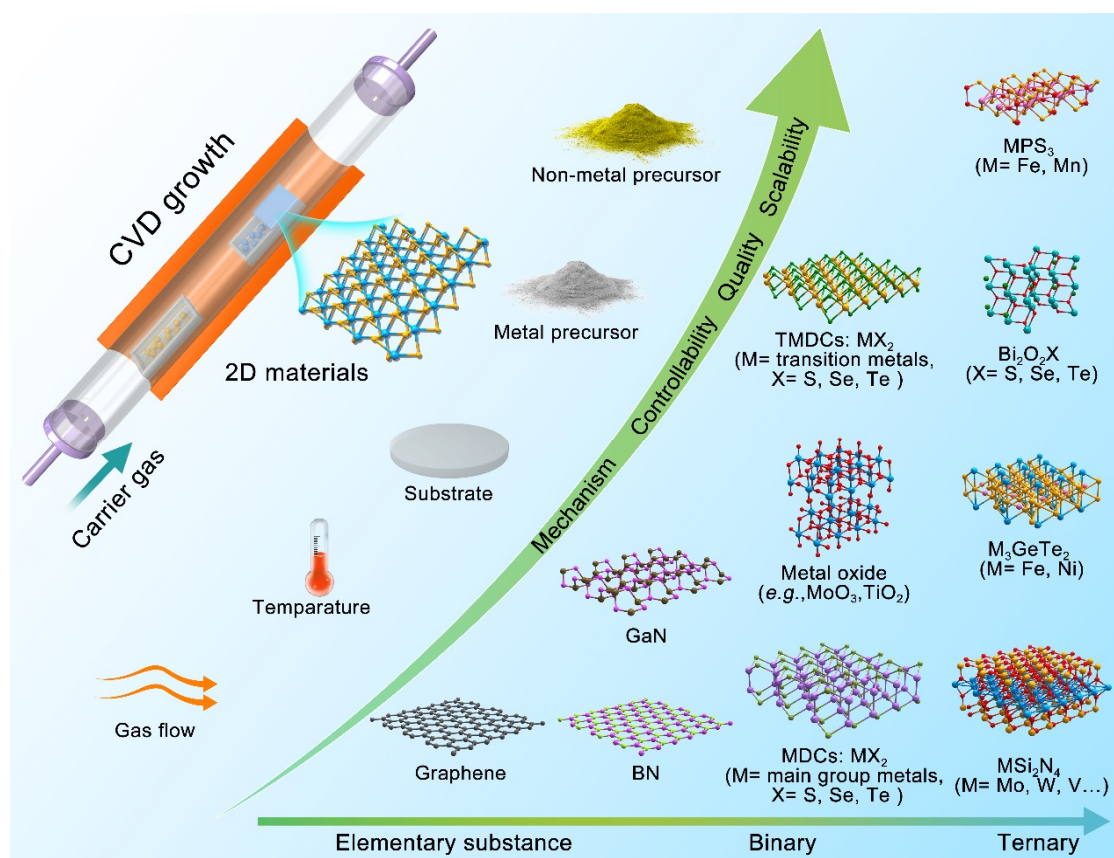

Figure 1. Schematic of the key parameters for the CVD growth of 2D materials ranging from elementary substance, binary, ternary to complicated materials.

2.1 Controlling the non-metal precursor

In general, solid-phase powder including non-metal precursors (*e.g.*, sulphur, selenium, and tellurium) and metal precursors (*e.g.*, molybdenum oxide, tungsten oxide) have been used as precursors to grow TMDCs in a traditional horizontal CVD (HCVD) system, and this suffers from intrinsic drawbacks of uncontrollable and nonuniform growth results (Figures 2a-c). It is caused by the time and mass-depended of the sublimation of solid precursors and spatially nonuniform growth dynamics which is known as the position-dependent phenomenon. Especially for the non-metal precursor, the sublimation process is difficult to control, and the local concentration is dependent on the position of substrate, so the as-grown samples are nonuniformly distributed on the substrate. Great efforts have been made to solve this issue by choosing metal-organic precursors[6] or changing the way of feeding the precursor into the system (*e.g.*, local feeding).[7]

Recently, instead of HCVD, our group designed a vertical CVD (VCVD) reactor and used gaseous precursors to grow monolayer centimeter-scale TMDCs with high uniformity (Figures 2d-f).[8] In this method, the gaseous precursors (*e.g.*, $H_2S$ and an Ar-bubbled metal precursor feed) substituted for the widely used solid precursors, so as to maintain a uniform precursor concentration gradient. The corresponding VCVD-grown TMDCs were characterized by optical microscopy (OM), Raman and photoluminescence (PL) spectra, showing the high uniformity (including morphology, nucleation density, and coverage), high quality over the centimeter-scale, and the average domain size of 9.7 μm. Compared with solid-precursor HCVD, the well-controlled gas flow, temperature, and precursor concentration gradients in VCVD produced more uniform growth dynamics, resulting in the controlled growth of monolayer TMDCs. Besides gaseous non-metal precursor, our group has also developed a liquid-precursor CVD (LCVD) method to prepare monolayer $MoS_2$,[9] where a $Na_2MoO_4$ solution was spin-coated on the substrate as the metal precursor and a sulphur-containing thiol dodecyl mercaptan ($C_{12}H_{25}SH$) solution was bubbled into the growth chamber by Ar as the non-metal precursor. This method provides the uniform supply of both precursors, leading to a homogeneous distribution of the $MoS_2$. Systematic microscopic and spectroscopic characterizations confirmed the good uniformity of the $MoS_2$. Overall, we believe that the method using gaseous or liquid non-metal precursors has better controllability than the ones using solid precursors. This strategy can be extended to Se- and Te-based TMDCs by using gaseous ($H_2Se$, $H_2Te$) or liquid precursors (Se-containing selenol and Te-containing telluromercapta), which will ensure a constant non-metal precursor feeding and improve the controllability of the CVD growth of TMDCs and other 2D compound materials.

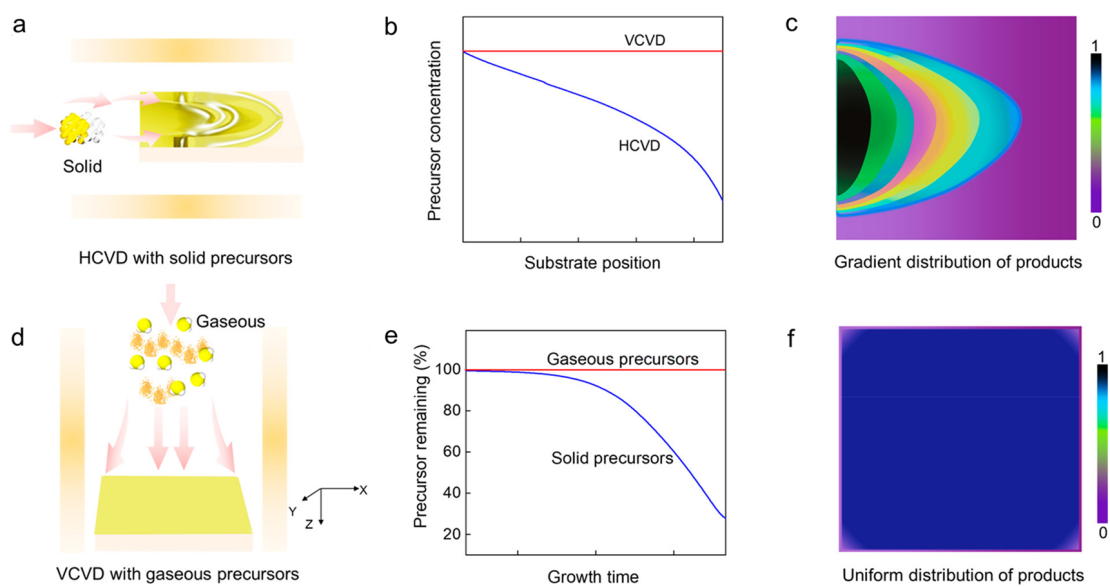

Figure 2. Comparison between solid-precursor HCVD and gaseous-precursor VCVD systems. (a, d) Schematics of the two systems. Relationships between (b) vapor concentration and substrate position and (e) remaining precursor and growth time in these two different CVD systems, respectively. (c, f) Growth profiles of these two different CVD systems. Reproduced with permission from ref 8. Copyright 2020 American Chemical Society.

## 2.2 Controlling the metal precursor

2D compound materials usually consist of non-metal and metal elements. In addition to the non-metal precursor, controlling the metal precursor is another major issue in the CVD growth of TMDCs. Solid powders like metal oxides or metal chlorides are usually used as metal precursors to grow TMDCs. Generally, the fast sublimation of the metal precursor leads to excessive nucleation sites and thick domains, while a slower sublimation rate will lead to uniform vapor concentration of metal precursor, which is essential for the growth of uniform monolayer TMDCs.[10] In recent years, great efforts have been made to solve this problem. For example, Lin et al. developed a metal film-guided growth method using a pre-deposited $MoO_3$ thin film as precursor in a sulphur vapor atmosphere to prepare $MoS_2$.[11] They obtained a uniform monolayer $MoS_2$ on inch-scale sapphire substrate. However, the domain size was small (less than 10 nm) and it was difficult to grow highly crystalline thin films. Kang et al. used a metal-organic CVD (MOCVD) system with gas-phase metal-organic compounds ($Mo(CO)_6$ and $W(CO)_6$) as the metal precursors, instead of the traditional metal oxide powders ($MoO_3$ and $WO_3$), and grew wafer-scale uniform TMDC films with a domain size less than 1 μm in 24 h.[6] This method gives better control of the metal precursor feeding during the growth process than traditional methods and indicates a way to grow TMDCs film with wafer-scale. Nevertheless, it is still a challenge to grow uniform 2D TMDC films with a large domain size, and the high toxicity of the metal-organic compounds is a drawback.

Porous molecular sieves with a high melting point have been also used to help grow 2D compound materials by adjusting the evaporation process of the metal precursor. For example, Shi et al. reported a method in which the Mo precursor was covered with an oxide inhibitor (*e.g.*, $SnO_2$, $Al_2O_3$) to maintain the gradual release of Mo concentration during CVD growth.[10] Consequently, the domain size, nucleation density, and number of layers of the as-grown $MoX_2$ (X = S, Se, Te) can be controlled by changing the amounts of the oxide inhibitor. This method has good universality and the use of oxide inhibitors could be extended to other molecular sieves, providing an effective method of controlling the nucleation and growth process of 2D compound materials.

Recently, our group reported a dissolution-precipitation (DP) growth method to control the metal precursor, and obtained uniform monolayer TMDCs over the whole centimeter-scale substrate (Figure 3a).[12] Different from the traditional CVD process (Figure 3c), in the DP growth method, the metal precursor was first buried in a confined space between a piece of top thin and another piece of bottom thick glass. The metal precursor then diffused to the surface of the upper glass where it served as a reactant to grow TMDCs at high temperature. The supply of the metal precursor in this method leads to a more uniform supply of metal source (Figure 3b). This strategy provided a simple way to control the metal precursor for the growth of TMDCs in a controlled manner and was expanded to grow different TMDCs and their alloys, such as $MoSe_2$, $WS_2$, $MoTe_2$, and $Mo_xW_{1-x}S_2$. In another recent work, Mo source diffused through a piece of top Cu substrate, and then reacted with nitrogen and silicon to form a new type of 2D compound material $MoSi_2N_4$.[13] We envision that by selecting suitable diffusion substrates, a series of 2D compound materials with widely tunable metal components could be controllably grown using a similar manner. Overall, the above results show that using a specific upper layer to confine the precursor in the sandwich-like structure is a promising way of controlling the sublimation and diffusion process of the metal source to obtain a suitable concentration for CVD growth. We believe that this strategy can be expanded to grow ternary (*e.g.*, $Bi_2O_2Se$) or even more complicated 2D compound materials.

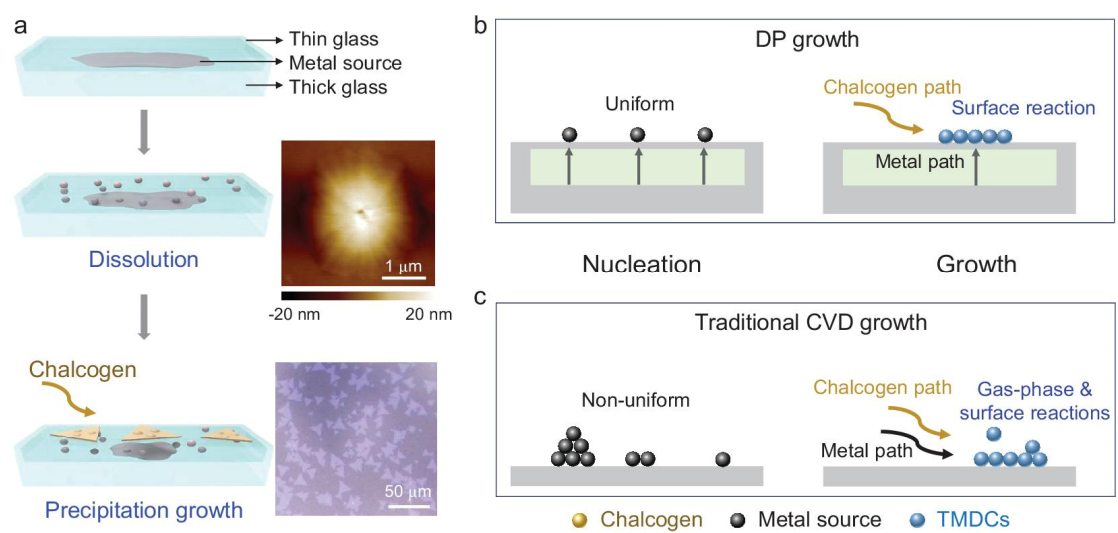

Figure 3. The DP growth of TMDCs in comparison with the traditional CVD method. (a) Illustration of the DP growth process. (b, c) Comparison between the DP growth and the traditional CVD growth of TMDCs in the nucleation and the growth processes. In DP growth, diffusion of metal and non-metal precursors is separated while in traditional CVD they share the same paths. Reproduced with permission from ref 12. Copyright 2020 Oxford University Press.

2.3 Substrate engineering

In terms of the controlled CVD growth of 2D materials, van der Waals (vdW)

epitaxial growth has proven to be an important technique on many substrates such as mica,[14] gallium nitride,[15] sapphire,[16] and quartz.[6] Thanks to the dangling bond-free surface of 2D materials, there are no strong chemical bonds between them and the substrate, so the lattice constant matching is not a must.[17] In recent years, great efforts have been made using substrate engineering to achieve a large domain size, fast growth rate, and consistent oriented growth of 2D compound materials, including selecting suitable substrates,[16] developing catalytic substrates,[18] and pre-treatment of the substrates (*e.g.*, annealing, selective patterning).[19, 20]

The first strategy is to select suitable substrates. Mica is a widely used substrate to grow 2D compound materials. For example, Ji et al. prepared monolayer $MoS_2$ on an inert and almost lattice-matched mica substrate using low-pressure CVD.[14] In addition to $MoS_2$, fluorophlogopite mica ($KMg_3AlSi_3O_{10}F_2$) is suitable for the epitaxial growth of other 2D compound materials, such as $In_2Se_3$,[21] $CrSe$,[22] and $Bi_2O_2Se$,[23] due to its atomic flatness, surface inertness, and high flexibility. Besides, sapphire is also one kind of substrate for the epitaxial growth of 2D materials. For example, Dumcenco et al. reported the epitaxy growth of high-quality monolayer $MoS_2$ on the sapphire substrate with controlling its lattice orientation.[16] Due to the epitaxial growth mechanism, the adjacent single islands formed a monolayer film as a consequent result. Overall, suitable substrates play important roles in the epitaxy growth of 2D materials.

The second strategy is to develop catalytically active substrates. Generally, one can decrease the active surface energy and thus will benefit the nucleation and growth of 2D materials during CVD growth.[24] However, for the growth of TMDCs, the inert non-metal substrates such as $SiO_2$/Si and sapphire are usually used, these substrates have poor catalytic activity, resulting in low growth rate and difficulty in controlling the growth behavior to produce a monolayer feature. Recently, Gao et al. reported the ultrafast catalytic growth of monolayer $WSe_2$ on Au foil with a growth rate of 26 μm s$^{-1}$.[18] This growth rate was 2-3 orders of magnitude higher than those of most monolayer TMDCs. As a result, millimeter-scale monolayer single-crystal $WSe_2$ domains were achieved in just 30 s and large continuous films in 60 s. Density functional theory (DFT) calculations showed that such a high growth rate of $WSe_2$ was caused by the small energy barriers for the diffusion of W and Se clusters on the Au substrate. This work shows that a surface with a low active surface energy like Au assists the nucleation of TMDCs during CVD process, providing a way to grow TMDCs at low temperature. However, because the most catalytic metals (*e.g.*, Cu, Ni) will react with the non-metal precursor (*e.g.*, sulphur), exploring the metal substrates with catalytic performance as well as inactivity to the non-metal precursor should be a promising way to achieve the growth of TMDCs at low temperature of < 400 °C.

The third strategy is pre-treatment of the substrates to obtain a specific dominant crystal plane for epitaxial growth of 2D materials, resulting in 2D domains with a consistent orientation. Substrates like $SiO_2$/Si do not have an epitaxial effect on 2D materials so that the as-grown TMDC domains are randomly oriented. As these single domains grow and merge, a large number of grain boundaries are formed between adjacent domains, leading to poor mechanical, electrical, and thermal properties.[25] Hence it is important to grow well-aligned TMDC single-crystal domains so that their seamless stitching results in single-crystal TMDCs with wafer-scale size. Several attempts have been made to solve this problem. For example, our group has developed a step-edge guided nucleation mechanism and achieved the aligned growth of WSe$_2$ using C-plane (0010) sapphire as substrate (Figure 4a).[26] After annealing at a high temperature (> 950 °C), the atomic steps formed on the sapphire surface served as active nucleation sites where well-aligned WSe$_2$ nuclei formed. With the growth time increasing, the WSe$_2$ tended to follow the layer-over-layer overlapping growth mode and finally formed aligned few-layer domains, as revealed by OM and scanning electron microscopy (Figures 4b and 4c). In a recent study, Aljarb et al. demonstrated the ledge-directed epitaxial growth of dense arrays of continuous, self-aligned, and monolayer single-crystal MoS$_2$ nanoribbons on β-gallium oxide (β-Ga$_2$O$_3$) (100) substrate (Figure 4d).[27] The MoS$_2$ seeds preferred to nucleate on the edges of exposed β-Ga$_2$O$_3$ (100) planes, and these aligned MoS$_2$ domains merged into centimeter-scale nanoribbons with an aspect ratio (length/width) of > 5000. This ledge-directed epitaxial growth mode also showed good universality for the growth of *p*-type WSe$_2$ nanoribbons and lateral heterostructures of *p*-type WSe$_2$ and *n*-type MoS$_2$. In addition, by combining the patterning technique to pre-treat the substrates, an array of 2D domains with designed shapes could be obtained. For example, Zhou et al. obtained selectively patterned nucleation sites on a mica substrate by using a polymethyl methacrylate lithography mask, which passivated the partial active sites and achieved the well-ordered growth of GaSe nanoplate arrays. The GaSe nanoplate arrays could be extended to triangular, hexagonal, round, and perforated morphologies (Figure 4e).[28] This method has been widely used to grow other arrays of 2D compound materials such as In$_2$Se$_3$ (Figure 4f)[29] and Bi$_2$O$_2$Se (Figure 4g).[30]

Overall, the vdW epitaxial growth of 2D compound materials on various substrates has made progress, but the growth of wafer-scale single-crystal 2D compound materials is still in its infancy. Designing suitable substrates with appropriate lattice symmetry may be an effective way to achieve this goal. In addition, traditional vdW epitaxy could be expanded to the liquid phase to grow the lateral perovskite heterostructures.[31]

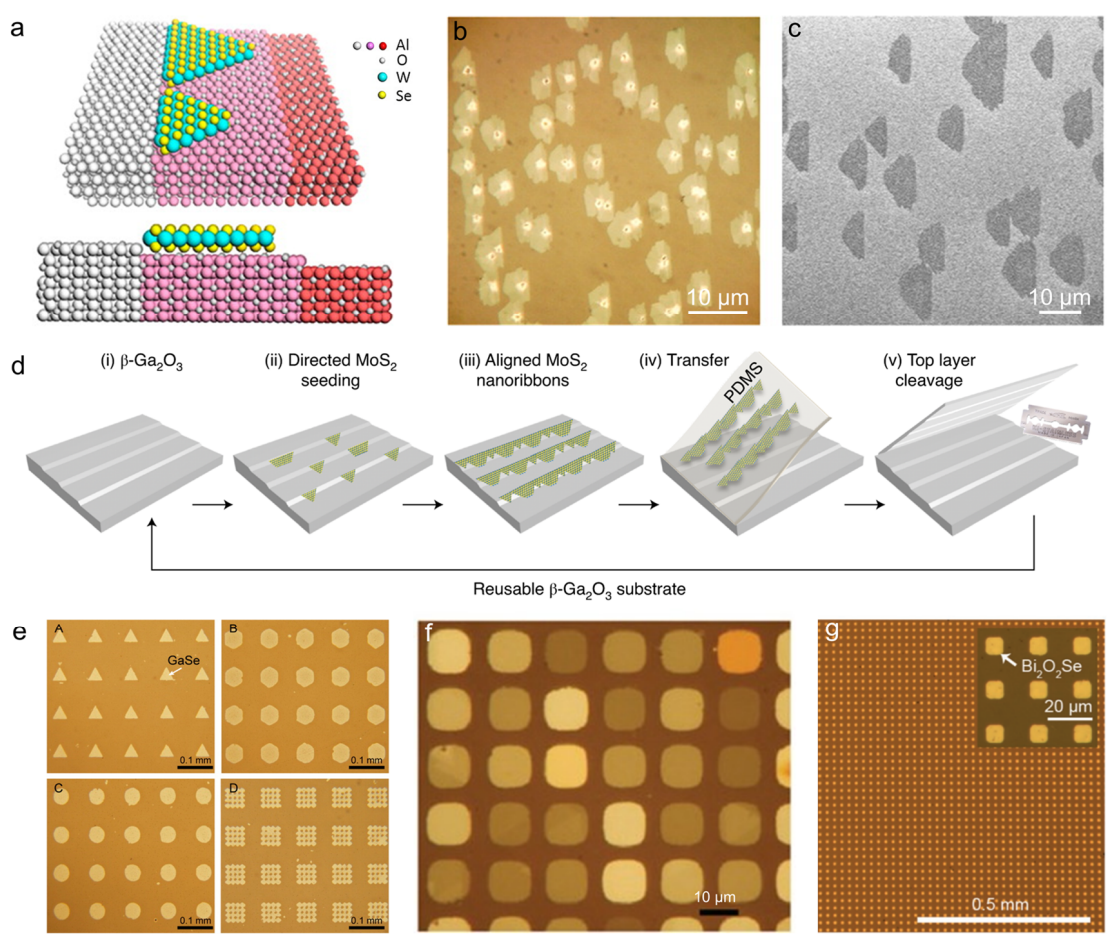

Figure 4. Substrate engineering for the aligned growth of 2D compound materials. (a) Schematic of the step-edge-guided nucleation and growth of aligned WSe$_2$ on a C-plane sapphire substrate and (b, c) its growth results. Reproduced with permission from ref 26. Copyright 2015 American Chemical Society. (d) Schematic of the sequential growth

of monolayer MoS$_2$ nanoribbons along the aligned ledges of a β-Ga$_2$O$_3$ (100) substrate. Reproduced with permission from ref 27. Copyright 2020 Springer Nature. (e) The selective-area epitaxial growth of 2D GaSe arrays and its typical OM images of triangular, hexagonal, round, and perforated shape. Reproduced with permission from ref 28. Copyright 2014 American Chemical Society. (f) 2D In$_2$Se$_3$ arrays obtained by selective-area epitaxial growth. Reproduced with permission from ref 29. Copyright 2015 Springer Nature. (g) 2D Bi$_2$O$_2$Se arrays obtained by selective-area epitaxial growth. Reproduced with permission from ref 30. Copyright 2017 Wiley-VCH.

2.4 Other parameters

In addition to the non-metal precursor, metal precursor and substrate engineering discussed above, there are some other parameters that are related to the CVD growth of 2D materials, such as the growth temperature and gas flow. Here, we discuss their influences in order to obtain suitable conditions for the controlled growth of 2D compound materials. Generally, the growth temperature affects the nucleation, domain size, number of layers, and crystallization of TMDCs during thermodynamically controlled growth processes. In the past, our group has made a comprehensive study of the influence of growth temperature on the growth of WSe$_2$ in ambient pressure CVD.[32] We obtained (i) WSe$_2$ particles at 800 °C, (ii) monolayer WSe$_2$ domains with a triangular shape at 950 °C, and (iii) WSe$_2$ with truncated triangular and hexagonal shapes at 1050 °C (Figures 5a and 5b). Meanwhile, the domain size and number of layers of WSe$_2$ increased with temperature in the range from 850 °C to 1050 °C. These results could be explained by the fact that the high growth temperature actuating the sublimation and diffusion of the $WO_3$ and $WO_{3-x}$ precursors, and further promoting the reactions between the $WO_3$, $WO_{3-x}$ and selenium precursors. But too fast a nucleation process would lead to multilayers or even bulk samples. This work shows a comprehensive understanding of the temperature effect on the growth of $WSe_2$ with feasible tunability. More attempts are still needed to investigate the growth process from a thermodynamic viewpoint. We can also simplify the process of investigating suitable reaction parameters with the help of machine learning and high throughput calculations. For example, Tang et al. reported a machine learning model to optimize the synthesis conditions of CVD-grown $MoS_2$, leading to the optimization of the experimental results with a minimized number of CVD trials.[33] Such combinations between computing technique with CVD growth process opens a promising way to accelerate the development of materials science.

For the growth of 2D materials with large domain size, the gas flow rate is crucial for control of the nucleation density, and has been studied in the growth of graphene[34] and BN.[35] Ideally, during vapor deposition the gas flow rate should be controlled to give viscous laminar flow in order to obtain a constant atmosphere. But in fact, there is always a velocity gradient of gases in the growth chamber, and the velocity decreases to zero near the substrate surface, forming a stagnant layer above it.[36] Since the precursor concentration on the substrate surface influences the nucleation density, its control is important for modulating the nucleation of the 2D material. Designing a micro-reactor with a confined space where the growth atmosphere is stable provides an alternative way to grow 2D materials with large domain size and high quality. In this regard, we prepared 2D In$_2$Se$_3$ in a confined microreactor made up of two face-to-face stacked mica substrates.[10] Because of the well-controlled precursor concentration, the as-prepared In$_2$Se$_3$ reached a size of over 200 μm. In addition, the development of new growth reactors that moderate the gas flow is a promising way to grow other 2D compounds. For example, Zhang et al. reported a general strategy for the growth of different heterostructures (*e.g.*, WS$_2$-WSe$_2$, WS$_2$-MoSe$_2$), multi-heterostructures (*e.g.*, WS$_2$-WSe$_2$-MoS$_2$, WS$_2$-MoSe$_2$-WSe$_2$), and superlattices (*e.g.*, WS$_2$-WSe$_2$-WS$_2$-WSe$_2$-WS$_2$) by precisely controlling the precursor supply with the assistance of a reverse gas flow design (Figures 5c-g).[37] In this design, the reverse flow not only terminates the uncontrollable nucleation between sequential vapor deposition growths by immediately blowing away the excess precursor, but also prevents the undesired thermal degradation from the exhaust heat by the reversed gas. This method improves the preparation of various 2D heterostructures and junctions by a continuous and sequential epitaxial growth mode, which may be also applied to other 2D complicated structures and 2D superlattices, such as perovskite junctions, and organic-inorganic hybrids.

Besides, the CVD growth microenvironment may also be changed by introducing additives such as oxygen,[38] fluorine,[39] and water vapor[40] into the reactions to modulate the growth chemistry of 2D materials. These strategies have shown advantages in obtaining the desired structure and morphology of 2D materials, including a large domain size and heterostructures. In addition, external fields such as laser, plasma, microwave, electric, and magnetic, can also be used to modulate the CVD growth process, for example, using a plasma-assisted CVD system to fabricate the Janus structure of TMDCs,[41] which may suggest designs for new CVD reactors for the growth of 2D compound materials with different structures and properties, especially for thermal-dynamically metastable materials.

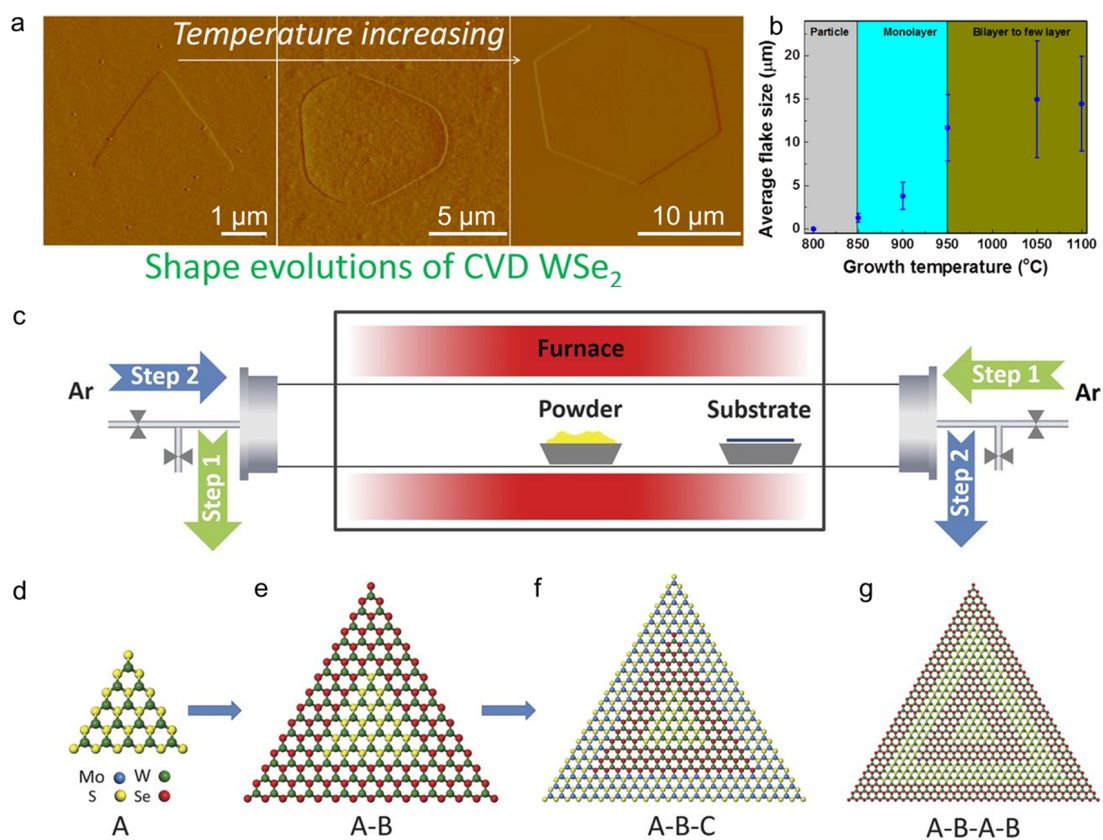

Figure 5. Effect of the temperature and gas flow on the growth of 2D compound materials. (a, b) Effect of the growth temperature on the domain size, number of layers, and morphology of CVD-grown $WSe_2$. Reproduced with permission from ref 32. Copyright 2015 American Chemical Society. (c) Schematic of a modified CVD system using reversed gas flow for the epitaxial growth of various 2D heterostructures. Growth results of (d) a monolayer seed A, (e) A-B heterostructures, (f) A-B-C multi-heterostructures, and (g) A-B-A-B superlattices. Reproduced with permission from ref 37. Copyright 2017 The American Association for the Advancement of Science.

2.5 Improving material quality

The performance of CVD grown-TMDCs based electronic and optoelectronic devices is greatly affected by the interfaces, defects, and grain boundaries in the materials, because they serve as carrier scattering centers and thus decrease the carrier mobility of the devices. Hence, the preparation of 2D compound materials with clean surfaces, high quality, and wafer-scale domain size should be a desired and urgent goal for future nanoelectronics.

To realize these goals, we first analyze the diffusion paths of non-metal and metal precursors in a traditional CVD system and find that they share the same diffusion path, resulting in unavoidable side reactions and the undesired deposition of particles or clusters on the surfaces of 2D materials. In contrast, our DP growth method described above overcomes this problem by separating the diffusion paths of the two precursors, thus obtaining $MoS_2$ with clean surface.[12] To characterize this cleanness, we exposed the monolayer $MoS_2$ to $TiCl_4$ vapor in humid air and found that there were only a few absorbed $TiO_2$ particles hydrolyzed from $TiCl_4$ on its surface (Figure 6b), rather than many $TiO_2$ particles absorbed on the traditional CVD grown $MoS_2$ (Figure 6a). We also checked the interaction of two DP-grown $MoS_2$ monolayers that had been manually-stacked and observed an interlayer emission peak (740 nm), further indicating the clean surfaces of the samples (Figure 6c). Field-effect transistors (FETs) based on DP-grown $MoS_2$ also showed a decent carrier mobility (7.5-21.5 $cm^2V^{-1}s^{-1}$) and a high on/off ratio ($10^6$-$10^8$) (Figure 6d). All these results confirmed the clean surface of the DP-grown

MoS$_2$ samples.

In addition, CVD-grown MoS$_2$ samples contain many sulphur vacancies. To minimize the density of these sulphur vacancies, our LCVD method used C$_{12}$H$_{25}$SH as precursor, which not only provided a continuous sulphur precursor but also *in-situ* repaired the sulphur vacancies during CVD growth.[9] In this way, we obtained MoS$_2$ samples with the lowest density of sulphur vacancies (0.32 nm$^{-2}$) among all CVD samples reported so far, as revealed by high-angle annular dark-field scanning transmission electron microscopy (HAADF-STEM) and the corresponding simulation result (Figures 6e and 6f). And the optical quality of the LCVD-grown MoS$_2$ was checked using PL measurements at low temperature (80 K) and showed that the full width at half maximum was 44 meV, which was closed to that of the exfoliated MoS$_2$ (Figure 6g). All these results prove the high quality of the LCVD-grown MoS$_2$. DFT calculations showed that using the dodecyl mercaptan as the sulphur precursor led to a relatively low energy barrier (0.61 eV and 0.19 eV) of the S-H bond breaking (Figure 6h). Therefore, the sulphur vacancies were repaired by using thiol during the *in-situ* growth process. The result is in good agreement with previous work of repairing the sulphur vacancies by post-treating defective MoS$_2$ with thiol molecules.[42] This work provides an effective route to grow MoS$_2$ with high quality, and the method can be extended to other 2D compound materials for use in high-performance devices.

Achieving wafer-scale single-crystal TMDC films has been a continuous pursuit because it has no grain boundaries, which is essential for high performance electronics and optoelectronics. Previous studies have reported that the epitaxial growth of TMDC films could be realized by stitching together different adjacent domains oriented at 0° and 60°,[16] but how to well control the nucleation and orientation in a wide range, and achieving the single-crystal TMDC films with wafer-scale by seamless stitching is still a big challenge. To this end, Yang et al. melted and re-solidified a commercial Au foil substrate to produce Au (111) surface and epitaxially grew a single-crystal, wafer-scale $MoS_2$ film (Figure 6i).[43] Using scanning tunneling microscopy combined with first-principle calculations, they showed that the nucleation of the monolayer $MoS_2$ was mainly guided by the steps on Au (111) surface. The highly oriented nucleation contributed to the growth of $MoS_2$ domains with the same orientation which merged into a large single crystal with one-inch size (Figure 6j). The FETs produced using such $MoS_2$ showed an average mobility (11.2 $cm^2V^{-1}s^{-1}$) and on/off ratio (7.7 ×$10^5$). This strategy may be extended for the growth of other 2D compound materials, facilitating their applications, especially in large area 2D nanoelectronics with high integration density.

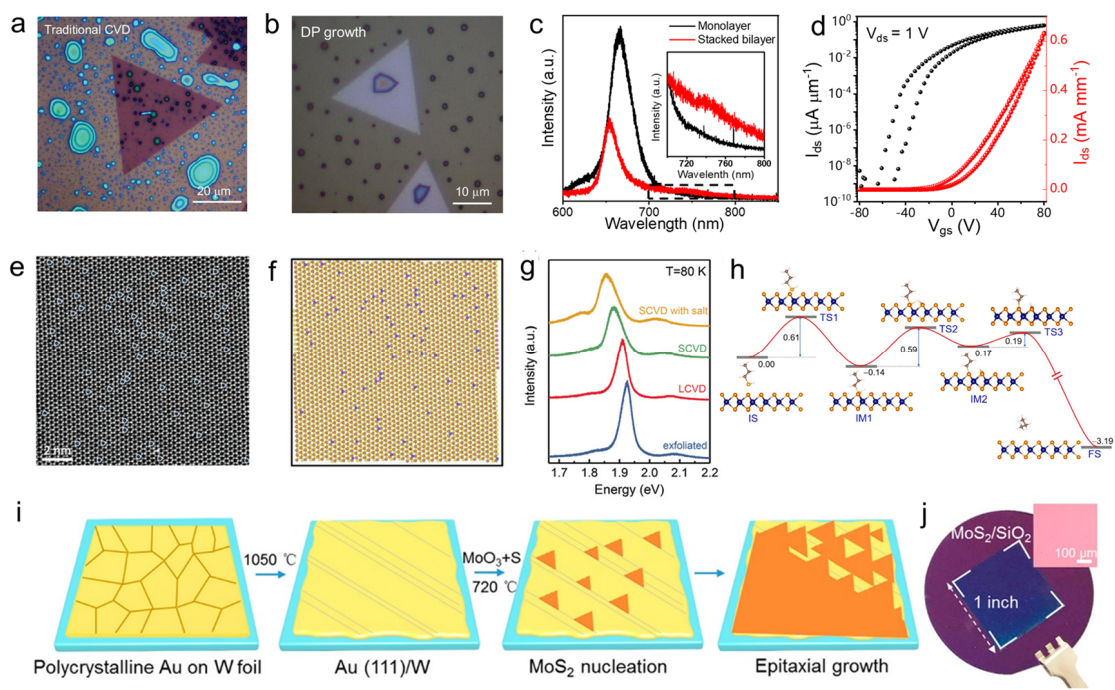

Figure 6. Urgent goals of the as-grown MoS$_2$ with clean surface, high quality, and large single-crystal. (a, b) Comparison of the surface cleanness between the traditional CVD-grown MoS$_2$ and the DP-grown MoS$_2$ after treatment with TiCl$_4$ vapor. (c) PL intensity of as-grown monolayer and manually-stacked bilayer DP-grown MoS$_2$. (d) Transfer curves of a FET based on the DP-grown MoS$_2$. Reproduced with permission from ref 12. Copyright 2020 Oxford University Press. (e, f) HAADF-STEM image and the simulation result of LCVD-grown MoS$_2$ by using thiol as precursor. The mono-sulphur vacancies are marked with blue circles. (g) Temperature-dependent PL spectra of MoS$_2$ grown by different methods. (h) DFT calculations of the reparation of sulphur vacancies by thiol molecules. Reproduced with permission from ref 9. Copyright 2020 Wiley-VCH. (i, j) Schematic of the formation of Au (111) steps and its effect on the growth of a single-crystal MoS$_2$ domain with 1-inch size. Reproduced with permission from ref 43. Copyright 2020 American Chemical Society.

2.6 Growth mechanism

It is difficult to obtain stable growth conditions using CVD because even small changes in the growth conditions may affect the reproducibility and controllability of the process. Hence, a deep understanding of the underlying growth mechanism is very important in the controlled growth of 2D compound materials. It is known that CVD involves typical heterogeneous nucleation reactions between the reactants occurring on the substrates. Heterogeneous nucleation usually needs a low surface energy, so how to modify it should be an important factor.[8] On one hand, we can change the compositions of the reactants to improve the sublimation and diffusion process.[44] On the other hand, we can also change the surface energy of the substrate to improve the wettability and thus assist its absorption of reactants.[24, 45]

Here, taking the growth of 2D $MoS_2$ as an example, when heating the two precursors ($MoO_3$ and sulphur powder) in a CVD system, their sublimation and diffusion rates play important roles in the domain size, thickness, and morphology of the as-grown $MoS_2$ domains. Usually, $MoO_3$ in the vapor phase undergoes the two-step reaction as shown below, where $MoO_3$ is reduced to be $MoO_{3-x}$ as the intermediate phase (Formula 1), which then diffuses onto the substrate by the carrier gas and reacts with sulphur vapor to grow 2D $MoS_2$ domains (Formula 2).

$$MoO_3 + x/2\ S \rightarrow MoO_{3-x} + x/2\ SO_2 \quad (1)$$

$$MoO_{3-x} + (7-x)/2\ S \rightarrow MoS_2 + (3-x)/2\ SO_2 \quad (2)$$

But the sublimation and the diffusion rates of the precursors are difficult to control due to the uncontrollable vapor concentrations when using solid precursors as mentioned above. Taking the fact that the high melting point of the metal oxide precursor results in a complicated growth process and limitations in the growth behaviors of the TMDCs (*e.g.*, position, yield, and domain size distribution), Zhou et al. reported that the salt additives (*e.g.*, NaCl, KCl) could react with metal oxides precursors (*e.g.*, $Nb_2O_5$, $MoO_3$, and $WO_3$) to transform the intermediate product from metal oxides into oxychloride compounds (*e.g.*, $NbO_xCl_y$, $MoO_xCl_y$, and $WO_xCl_y$) during the growth process.[44] Compared to the reactions between metal oxide and sulphur, the salt additives not only decrease the melting points of these reactants in the growth system, but also facilitate the easy sublimation of the metal precursors and thus accelerate the nucleation and reaction rate. Here, the growth steps are generally shown below:

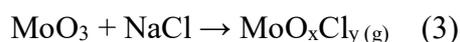

$$MoO_3 + NaCl \rightarrow MoO_xCl_{y\,(g)} \quad (3)$$

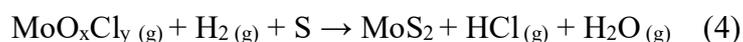

$$MoO_xCl_{y\,(g)} + H_{2\,(g)} + S \rightarrow MoS_2 + HCl_{(g)} + H_2O_{(g)} \quad (4)$$

Based on this design, they prepared a range of 2D TMDCs, including 32 binary compounds based on the transition metals (Ti, Zr, Hf, V, Nb, Ta, Mo, W, Re, Pt, Pd, and Fe), 13 alloys (11 ternary, one quaternary, and one quinary), as well as two heterostructured compounds (Figure 7a). This work provides a universal growth method for the growth of 2D TMDCs,[46] especially for metallic TMDCs which are hard to grow due to the high melting points of the corresponding metal precursors.

Besides modulating growth chemistry by additives, substrate modification can also affect the growth of 2D materials. As mentioned before, the ability of the substrate to absorb the precursors and thus assist the nucleation is highly related to its surface energy. Generally, heterogeneous nucleation occurs at the preferential sites with a low surface energy. To modulate the surface energy of the substrate, great efforts have been made such as using aromatic molecules (*e.g.*, perylene-3,4,9,10-tetracarboxylic acid tetrapotassium salt (PTAS), $F_{16}CuPc$) as seeding promoters[24] and pre-patterning the seeds on the substrate.[45] These methods lead to a controlled growth behavior (*e.g.*, uniform nucleation sites, growth, continuous thin film, and large single-crystal thin films). For example, Ling et al. reported that using PTAS as a seeding promoter help grow uniform large-area, highly crystalline $MoS_2$ at a relatively low growth temperature (650 °C) as shown in Figure 7b.[24] They further proved that an optimized dosage of seeding molecules could assist the nucleation of the $MoS_2$ domains. As for the underlying mechanism, the promoter can reduce the surface energy of the substrate, which increases its wettability for subsequent nucleation of $MoS_2$. However, with only the additive as a seeding promoter, it is difficult to precisely control the orientation and morphology of the TMDCs, and the mechanism of seeding nucleation still needs to be proved. Recently, Li et al. used Au nanoparticles as seeds instead of PTAS to grow TMDCs (Figure 7c).[45] After combining it with a photolithography technique, they obtained pre-patterned Au nanostructure arrays. Based on the Au-seeded substrate, monolayer $MoS_2$ nucleated and grew on the exposed Au area, forming precisely defined geometries due to the good affinity between $MoS_2$ and the Au nanoparticles. DFT calculations revealed the seeding mechanism of the Au nanoparticles for the nucleation of $MoS_2$ which contained two steps. First, an initial few-layer $MoS_2$ shell was formed on the surface of Au nanoparticles. Second, $MoS_2$ fully covered the Au nanoparticles and subsequently grew along the $SiO_2$/Si substrate. This work serves as an important step in understanding the fundamental mechanism of the influence of a specific substrate on the nucleation of $MoS_2$. This kind of substrate engineering also indicates the potential to grow other 2D compound materials with designed patterns or arrays.

Overall, from these typical cases mentioned above, the nucleation and the growth of TMDCs are inseparably intertwined. Therefore, effective methods that can control the nucleation also can be used to modulate the growth behavior of 2D compound materials. More works are needed to understand how to modulate the interfaces between reactants and substrates, *e.g.*, decreasing the surface energy barriers by using catalysts, reducing the growth temperature under high energy inputs like a plasma, microwave.

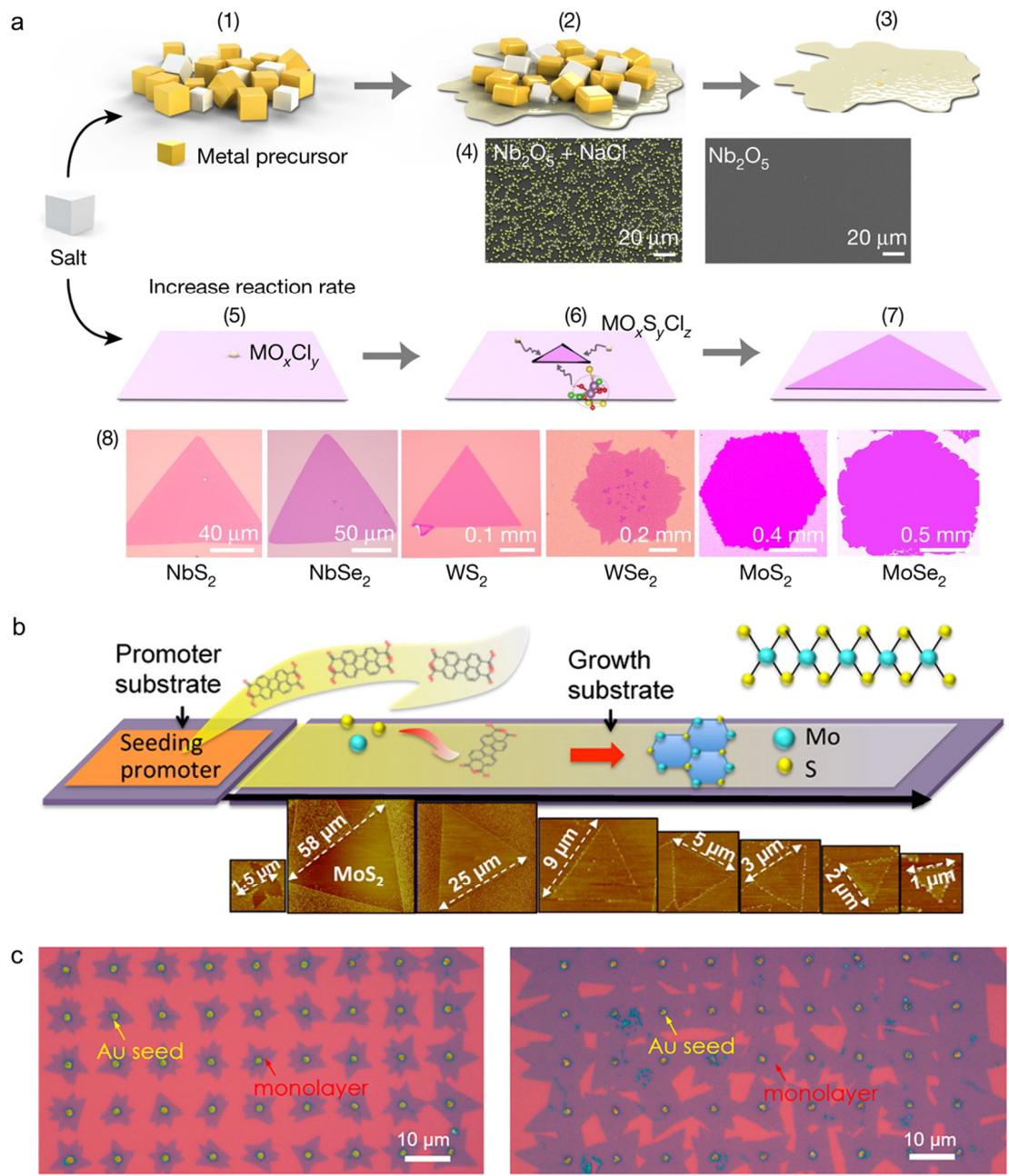

Figure 7. The nucleation mechanism for the growth of TMDCs is clarified by self-seeding and substrate surface-seeding methods. (a) Reaction mechanism of the salt-assisted growth of 2D TMDCs. Reproduced with permission from ref 44. Copyright 2018 Springer Nature. (b) Schematic of the role of a seeding promoter in the growth of MoS$_2$ by CVD. Reproduced with permission from ref 24. Copyright 2014 American Chemical Society. (c) Au-seeded CVD growth of MoS$_2$ arrays. Reproduced with permission from ref 45. Copyright 2018 American Chemical Society.

## 3. CONCLUSIONS AND OUTLOOK

In this Account, we have summarized the state-of-the-art of CVD-grown 2D compound materials using TMDCs as examples. We first reviewed the key parameters affecting the CVD growth process, which can be tuned to improve the controllability of the growth of 2D TMDCs, including the non-metal and metal precursors, substrate engineering, temperature, and gas flow. Then, the current attends on improving material quality and mechanism understanding are discussed. Despite its rapid progress, the development of 2D compound materials like TMDCs is still in its early stages. A large number of other 2D compound materials have not been well studied or have not been successfully grown yet, and more exciting discoveries are waiting to be found. In the following text, we propose a few potential research directions in this emerging field.

**Exploration and growth of novel 2D compound materials.** With the assistance of high-throughput computation based on the materials database, it is facile to find more 2D compound materials with unusual structures and properties.[47] In addition, compared with the large numbers of *n*-type 2D materials, the number of *p*-type materials is limited and more effort is needed to seek stable *p*-type 2D materials. For the preparation of other novel 2D compound materials, it is feasible to grow materials with wide bandgap (> 3.0 eV) or narrow bandgap (< 0.3 eV). Another promising research direction is the direct growth of metastable materials. Taking black phosphorus as an example, it exhibits a thickness-dependent direct bandgap from 0.3 eV to 1.5 eV, and how to prepare high-quality black phosphorus with atomic-thickness deserves further study.

**Using 2D compound materials as platforms to grow heterostructures.** A family of heterostructures may be grown by using 2D materials as templates or platforms, for example, exposing the existed 2D materials to CVD setups with plasma, laser, microwave, infrared, or rapid heat annealing conditions, which may produce many novel structures, such as metal single atoms, non-metal single atoms, amorphous structures, nanopores, Janus structures. In addition, the two-step CVD method may form new structures. Such treatments include boration, carbonation, nitridation, oxidation, phosphating, sulfuration, hydrodesulfurization, and reaction with n-butyl lithium. Noteworthily, the use of 2D materials as substrates for the vdW epitaxial growth of macroscale and high-quality "magic-angle 2D structures" with special and various twist angles will be an interesting topic.

**Exploration of new physics and applications of using 2D compound materials.** 2D materials with exotic properties like room-temperature ferromagnetism or high-temperature superconductivity are desired in spintronics and electronics. Besides, 2D materials with wide bandgap and narrow bandgap will provide more candidates for the blue light-emitting displayers and far- and mid-infrared light detection, respectively. 2D membranes with ion-selective channels or pores provide high energy extraction efficiency in ion transport, therefore, the controlled growth of nanoporous 2D films with a well-distributed pore size is important to provide a high power density and energy conversion efficiency for molecule and ion separation, transport, and energy devices.

AUTHOR INFORMATION

Corresponding Authors

*E-mail: bilu.liu@sz.tsinghua.edu.cn

*E-mail: hmcheng@sz.tsinghua.edu.cn

ORCID

Bilu Liu: 0000-0002-7274-5752

Hui-Ming Cheng: 0000-0002-5387-4241

Notes

The authors declare no competing financial interest.

Biographies

Lei Tang received his master's degree from Shanghai University and Suzhou Institute of Nano-Tech and Nano-Bionics, Chinese Academy of Sciences (2016), under the supervision of Prof. Yagang Yao. Currently, he is a Ph.D. candidate under the supervision of Profs. Bilu Liu and Hui-Ming Cheng in Tsinghua-Berkeley Shenzhen Institute, Tsinghua University. His research interests are focused on the controlled growth of 2D materials and their electronic and optoelectronic properties.

Junyang Tan received his bachelor's degree in the School of Materials Science and Engineering at Northeastern University in 2018. Currently, he is a Ph.D. student under the supervision of Profs. Bilu Liu and Hui-Ming Cheng in Tsinghua-Berkeley Shenzhen Institute at Tsinghua University. His current research interests are mainly focused on the controlled preparation of 2D materials and their heterostructures.

Huiyu Nong received her bachelor's degree in Materials Science and Engineering from Tsinghua University in 2019. She is currently pursuing a Ph.D. degree in Tsinghua-Berkley Shenzhen Institute, Tsinghua University. Her research interests are focused on the synthesis and physical properties of 2D materials.

Bilu Liu is currently an Associate Professor and a principal investigator at Tsinghua-Berkeley Shenzhen Institute (TBSI), Tsinghua University, China. His research interests cover the chemistry and materials science of low-dimensional materials with an emphasis on carbon nanostructures, 2D materials, and their heterostructures. His work relates to the controlled preparation of these materials and their applications in electronics, optoelectronics, and catalysis.

Hui-Ming Cheng is currently a professor of the Advanced Carbon Research Division at Shenyang National Laboratory for Materials Science, IMR, CAS, and also the Low-Dimensional Materials and Devices Laboratory at Tsinghua-Berkeley Shenzhen Institute (TBSI), Tsinghua University, China. His research interests are focused on the synthesis and applications of carbon nanotubes, graphene, other 2D materials, and high-performance bulk carbons, and on the development of new energy materials for batteries, electrochemical capacitors, and hydrogen production from water by photocatalysis.

## ACKNOWLEDGMENTS

We acknowledge support from the National Natural Science Foundation of China (Nos. 51722206, 51991340, 51991343, and 51920105002,), the Youth 1000-Talent Program of China, the National Key R&D Program (2018YFA0307200), Guangdong Innovative and Entrepreneurial Research Team Program (No. 2017ZT07C341), the Bureau of Industry and Information Technology of Shenzhen for the "2017 Graphene Manufacturing Innovation Center Project" (No. 201901171523).

## REFERENCES

1. Novoselov, K. S.; Geim, A. K.; Morozov, S. V.; Jiang, D.; Zhang, Y.; Dubonos, S. V.; Grigorieva, I. V.; Firsov, A. A., Electric field effect in atomically thin carbon films. *Science* **2004,** *306* (5696), 666.

2. Sangwan, V. K.; Lee, H. S.; Bergeron, H.; Balla, I.; Beck, M. E.; Chen, K. S.; Hersam, M. C., Multi-terminal memtransistors from polycrystalline monolayer molybdenum disulfide. *Nature* **2018,** *554* (7693), 500-504.

3. Yin, Z.; Li, H.; Li, H.; Jiang, L.; Shi, Y.; Sun, Y.; Lu, G.; Zhang, Q.; Chen, X.;

Zhang, H., Single-layer MoS$_2$ phototransistors. *ACS Nano* **2012,** *6* (1), 74-80.

4. Li, N.; Wang, Q.; Shen, C.; Wei, Z.; Yu, H.; Zhao, J.; Lu, X.; Wang, G.; He, C.; Xie, L.; Zhu, J.; Du, L.; Yang, R.; Shi, D.; Zhang, G., Large-scale flexible and transparent electronics based on monolayer molybdenum disulfide field-effect transistors. *Nat Electron* **2020**.

5. Zhang, C.; Tan, J. Y.; Pan, Y. K.; Cai, X. K.; Zou, X. L.; Cheng, H. M.; Liu, B. L., Mass production of 2D materials by intermediate-assisted grinding exfoliation. *Natl Sci Rev* **2020,** *7* (2), 324-332.

6. Kang, K.; Xie, S.; Huang, L.; Han, Y.; Huang, P. Y.; Mak, K. F.; Kim, C. J.; Muller, D.; Park, J., High-mobility three-atom-thick semiconducting films with wafer-scale homogeneity. *Nature* **2015,** *520* (7549), 656-60.

7. He, T.; Li, Y.; Zhou, Z.; Zeng, C.; Qiao, L.; Lan, C.; Yin, Y.; Li, C.; Liu, Y., Synthesis of large-area uniform MoS$_2$ films by substrate-moving atmospheric pressure chemical vapor deposition: from monolayer to multilayer. *2D Materials* **2019,** *6* (2).

8. Tang, L.; Li, T.; Luo, Y.; Feng, S.; Cai, Z.; Zhang, H.; Liu, B.; Cheng, H.-M., Vertical chemical vapor deposition growth of highly uniform 2D transition metal dichalcogenides. *ACS Nano* **2020,** *14* (4), 4646-4653.

9. Feng, S.; Tan, J.; Zhao, S.; Zhang, S.; Khan, U.; Tang, L.; Zou, X.; Lin, J.; Cheng, H. M.; Liu, B., Synthesis of ultrahigh-quality monolayer molybdenum disulfide through in situ defect healing with thiol molecules. *Small* **2020**, e2003357.

10. Shi, R.; He, P.; Cai, X.; Zhang, Z.; Wang, W.; Wang, J.; Feng, X.; Wu, Z.; Amini, A.; Wang, N.; Cheng, C., Oxide inhibitor-assisted growth of single-layer molybdenum dichalcogenides (MoX$_2$, X = S, Se, Te) with controllable molybdenum release. *ACS Nano* **2020,** *14* (6), 7593-7601.

11. Lin, Y.-C.; Zhang, W.; Huang, J.-K.; Liu, K.-K.; Lee, Y.-H.; Liang, C.-T.; Chu, C.-W.; Li, L.-J., Wafer-scale MoS$_2$ thin layers prepared by MoO$_3$ sulfurization. *Nanoscale* **2012,** *4* (20), 6637-6641.

12. Cai, Z.; Lai, Y.; Zhao, S.; Zhang, R.; Tan, J.; Feng, S.; Zou, J.; Tang, L.; Lin, J.; Liu, B.; Cheng, H.-M., Dissolution-precipitation growth of uniform and clean two dimensional transition metal dichalcogenides. *Natl Sci Rev* **2020**.

13. Hong, Y. L.; Liu, Z.; Wang, L.; Zhou, T.; Ma, W.; Xu, C.; Feng, S.; Chen, L.; Chen, M. L.; Sun, D. M.; Chen, X. Q.; Cheng, H. M.; Ren, W., Chemical vapor deposition of layered two-dimensional MoSi$_2$N$_4$ materials. *Science* **2020,** *369* (6504), 670-674.

14. Ji, Q.; Zhang, Y.; Gao, T.; Zhang, Y.; Ma, D.; Liu, M.; Chen, Y.; Qiao, X.; Tan, P.-H.; Kan, M.; Feng, J.; Sun, Q.; Liu, Z., Epitaxial monolayer MoS$_2$ on mica with novel photoluminescence. *Nano Lett* **2013,** *13* (8), 3870-3877.

15. Ruzmetov, D.; Zhang, K.; Stan, G.; Kalanyan, B.; Bhimanapati, G. R.; Eichfeld, S. M.; Burke, R. A.; Shah, P. B.; O'Regan, T. P.; Crowne, F. J.; Birdwell, A. G.; Robinson, J. A.; Davydov, A. V.; Ivanov, T. G., Vertical 2D/3D semiconductor heterostructures based on epitaxial molybdenum disulfide and gallium nitride. *ACS Nano* **2016,** *10* (3), 3580-8.

16. Dumcenco, D.; Ovchinnikov, D.; Marinov, K.; Lazic, P.; Gibertini, M.; Marzari, N.; Lopez Sanchez, O.; Kung, Y. C.; Krasnozhon, D.; Chen, M. W.; Bertolazzi, S.; Gillet, P.; Fontcuberta i Morral, A.; Radenovic, A.; Kis, A., Large-area epitaxial monolayer

MoS$_2$. *ACS Nano* **2015,** *9* (4), 4611-20.

17. Wang, M.; Wu, J.; Lin, L.; Liu, Y.; Deng, B.; Guo, Y.; Lin, Y.; Xie, T.; Dang, W.; Zhou, Y.; Peng, H., Chemically engineered substrates for patternable growth of two-dimensional chalcogenide crystals. *ACS Nano* **2016,** *10* (11), 10317-10323.

18. Gao, Y.; Hong, Y. L.; Yin, L. C.; Wu, Z.; Yang, Z.; Chen, M. L.; Liu, Z.; Ma, T.; Sun, D. M.; Ni, Z.; Ma, X. L.; Cheng, H. M.; Ren, W., Ultrafast growth of high-quality monolayer WSe$_2$ on Au. *Adv Mater* **2017,** *29* (29), 1700990.

19. Yu, H.; Liao, M.; Zhao, W.; Liu, G.; Zhou, X. J.; Wei, Z.; Xu, X.; Liu, K.; Hu, Z.; Deng, K.; Zhou, S.; Shi, J. A.; Gu, L.; Shen, C.; Zhang, T.; Du, L.; Xie, L.; Zhu, J.; Chen, W.; Yang, R.; Shi, D.; Zhang, G., Wafer-scale growth and transfer of highly-oriented monolayer MoS$_2$ continuous films. *ACS Nano* **2017,** *11* (12), 12001-12007.

20. Li, J.; Yang, X.; Liu, Y.; Huang, B.; Wu, R.; Zhang, Z.; Zhao, B.; Ma, H.; Dang, W.; Wei, Z.; Wang, K.; Lin, Z.; Yan, X.; Sun, M.; Li, B.; Pan, X.; Luo, J.; Zhang, G.; Liu, Y.; Huang, Y.; Duan, X.; Duan, X., General synthesis of two-dimensional van der Waals heterostructure arrays. *Nature* **2020**.

21. Tang, L.; Teng, C.; Luo, Y.; Khan, U.; Pan, H.; Cai, Z.; Zhao, Y.; Liu, B.; Cheng, H.-M., Confined van der Waals epitaxial growth of two-dimensional large single-crystal In$_2$Se$_3$ for flexible broadband photodetectors. *Research* **2019,** *2019*, 1-10.

22. Zhang, Y.; Chu, J.; Yin, L.; Shifa, T. A.; Cheng, Z.; Cheng, R.; Wang, F.; Wen, Y.; Zhan, X.; Wang, Z.; He, J., Ultrathin magnetic 2D single-crystal CrSe. *Adv Mater* **2019,** *31* (19), e1900056.

23. Khan, U.; Luo, Y.; Tang, L.; Teng, C.; Liu, J.; Liu, B.; Cheng, H. M., Controlled

vapor–solid deposition of millimeter-size single crystal 2D $Bi_2O_2Se$ for high-performance phototransistors. *Adv Funct Mater* **2019,** *29* (14).

24. Ling, X.; Lee, Y. H.; Lin, Y.; Fang, W.; Yu, L.; Dresselhaus, M. S.; Kong, J., Role of the seeding promoter in $MoS_2$ growth by chemical vapor deposition. *Nano Lett* **2014,** *14* (2), 464-72.

25. Zhang, J.; Lin, L.; Jia, K.; Sun, L.; Peng, H.; Liu, Z., Controlled growth of single-crystal graphene films. *Adv Mater* **2020,** *32* (1), e1903266.

26. Chen, L.; Liu, B.; Ge, M.; Ma, Y.; Abbas, A. N.; Zhou, C., Step-edge-guided nucleation and growth of aligned $WSe_2$ on sapphire via a layer-over-layer growth mode. *ACS Nano* **2015,** *9* (8), 8368-75.

27. Aljarb, A.; Fu, J. H.; Hsu, C. C.; Chuu, C. P.; Wan, Y.; Hakami, M.; Naphade, D. R.; Yengel, E.; Lee, C. J.; Brems, S.; Chen, T. A.; Li, M. Y.; Bae, S. H.; Hsu, W. T.; Cao, Z.; Albaridy, R.; Lopatin, S.; Chang, W. H.; Anthopoulos, T. D.; Kim, J.; Li, L. J.; Tung, V., Ledge-directed epitaxy of continuously self-aligned single-crystalline nanoribbons of transition metal dichalcogenides. *Nat Mater* **2020**.

28. Zhou, Y.; Nie, Y.; Liu, Y.; Yan, K.; Hong, J.; Jin, C.; Zhou, Y.; Yin, J.; Liu, Z.; Peng, H., Epitaxy and photoresponse of two-dimensional GaSe crystals on flexible transparent mica sheets. *ACS Nano* **2014,** *8* (2), 1485-90.

29. Zheng, W.; Xie, T.; Zhou, Y.; Chen, Y. L.; Jiang, W.; Zhao, S.; Wu, J.; Jing, Y.; Wu, Y.; Chen, G.; Guo, Y.; Yin, J.; Huang, S.; Xu, H. Q.; Liu, Z.; Peng, H., Patterning two-dimensional chalcogenide crystals of $Bi_2Se_3$ and $In_2Se_3$ and efficient photodetectors. *Nat Commun* **2015,** *6*, 6972.

30. Wu, J.; Liu, Y.; Tan, Z.; Tan, C.; Yin, J.; Li, T.; Tu, T.; Peng, H., Chemical patterning of high-mobility semiconducting 2D Bi$_2$O$_2$Se crystals for integrated optoelectronic devices. *Adv Mater* **2017,** *29* (44), 1704060.

31. Shi, E.; Yuan, B.; Shiring, S. B.; Gao, Y.; Akriti; Guo, Y.; Su, C.; Lai, M.; Yang, P.; Kong, J.; Savoie, B. M.; Yu, Y.; Dou, L., Two-dimensional halide perovskite lateral epitaxial heterostructures. *Nature* **2020,** *580* (7805), 614-620.

32. Liu, B.; Fathi, M.; Chen, L.; Abbas, A.; Ma, Y.; Zhou, C., Chemical vapor deposition growth of monolayer WSe$_2$ with tunable device characteristics and growth mechanism study. *ACS Nano* **2015,** *9* (6), 6119-27.

33. Tang, B., ; Lu, Yu,; Zhou, J,; Wang, H,; Golani, P,; Xu, Mna,; Xu, Q.; Guan, C,; Liu, Z., Machine learning-guided synthesis of advanced inorganic materials. *arXiv:1905.03938* **2020**.

34. Li, X.; Magnuson, C. W.; Venugopal, A.; Tromp, R. M.; Hannon, J. B.; Vogel, E. M.; Colombo, L.; Ruoff, R. S., Large-area graphene single crystals grown by low-pressure chemical vapor deposition of methane on copper. *J Am Chem Soc* **2011,** *133* (9), 2816-2819.

35. Liu, J.; Yu, L.; Cai, X.; Khan, U.; Cai, Z.; Xi, J.; Liu, B.; Kang, F., Sandwiching h-BN monolayer films between sulfonated poly(ether ether ketone) and nafion for proton exchange membranes with improved ion selectivity. *ACS Nano* **2019,** *13* (2), 2094-2102.

36. Bhaviripudi, S.; Jia, X.; Dresselhaus, M. S.; Kong, J., Role of Kinetic Factors in Chemical Vapor Deposition Synthesis of uniform large area graphene using copper catalyst. *Nano Lett* **2010,** *10* (10), 4128-4133.

37. Zhang, Z.; Chen, P.; Duan, X.; Zang, K.; Luo, J.; Duan, X., Robust epitaxial growth of two-dimensional heterostructures, multiheterostructures, and superlattices. *Science* **2017,** *357* (6353), 788-792.

38. Chen, W.; Zhao, J.; Zhang, J.; Gu, L.; Yang, Z.; Li, X.; Yu, H.; Zhu, X.; Yang, R.; Shi, D.; Lin, X.; Guo, J.; Bai, X.; Zhang, G., Oxygen-assisted chemical vapor deposition growth of large single-crystal and high-quality monolayer $MoS_2$. *J Am Chem Soc* **2015,** *137* (50), 15632-5.

39. Liu, C.; Xu, X.; Qiu, L.; Wu, M.; Qiao, R.; Wang, L.; Wang, J.; Niu, J.; Liang, J.; Zhou, X.; Zhang, Z.; Peng, M.; Gao, P.; Wang, W.; Bai, X.; Ma, D.; Jiang, Y.; Wu, X.; Yu, D.; Wang, E.; Xiong, J.; Ding, F.; Liu, K., Kinetic modulation of graphene growth by fluorine through spatially confined decomposition of metal fluorides. *Nat Chem* **2019,** *11* (8), 730-736.

40. Sahoo, P. K.; Memaran, S.; Xin, Y.; Balicas, L.; Gutierrez, H. R., One-pot growth of two-dimensional lateral heterostructures via sequential edge-epitaxy. *Nature* **2018,** *553* (7686), 63-67.

41. Lu, A. Y.; Zhu, H.; Xiao, J.; Chuu, C. P.; Han, Y.; Chiu, M. H.; Cheng, C. C.; Yang, C. W.; Wei, K. H.; Yang, Y.; Wang, Y.; Sokaras, D.; Nordlund, D.; Yang, P.; Muller, D. A.; Chou, M. Y.; Zhang, X.; Li, L. J., Janus monolayers of transition metal dichalcogenides. *Nat Nanotechnol* **2017,** *12* (8), 744-749.

42. Yu, Z.; Pan, Y.; Shen, Y.; Wang, Z.; Ong, Z. Y.; Xu, T.; Xin, R.; Pan, L.; Wang, B.; Sun, L.; Wang, J.; Zhang, G.; Zhang, Y. W.; Shi, Y.; Wang, X., Towards intrinsic charge transport in monolayer molybdenum disulfide by defect and interface engineering. *Nat*

*Commun* **2014,** *5*, 5290.

43. Yang, P.; Zhang, S.; Pan, S.; Tang, B.; Liang, Y.; Zhao, X.; Zhang, Z.; Shi, J.; Huan, Y.; Shi, Y.; Pennycook, S. J.; Ren, Z.; Zhang, G.; Chen, Q.; Zou, X.; Liu, Z.; Zhang, Y., Epitaxial growth of centimeter-scale single-crystal MoS$_2$ monolayer on Au (111). *ACS Nano* **2020,** *14* (4), 5036-5045.

44. Zhou, J.; Lin, J.; Huang, X.; Zhou, Y.; Chen, Y.; Xia, J.; Wang, H.; Xie, Y.; Yu, H.; Lei, J.; Wu, D.; Liu, F.; Fu, Q.; Zeng, Q.; Hsu, C. H.; Yang, C.; Lu, L.; Yu, T.; Shen, Z.; Lin, H.; Yakobson, B. I.; Liu, Q.; Suenaga, K.; Liu, G.; Liu, Z., A library of atomically thin metal chalcogenides. *Nature* **2018,** *556* (7701), 355-359.

45. Li, Y.; Hao, S.; DiStefano, J. G.; Murthy, A. A.; Hanson, E. D.; Xu, Y.; Wolverton, C.; Chen, X.; Dravid, V. P., Site-specific positioning and patterning of MoS$_2$ monolayers: the role of Au seeding. *ACS Nano* **2018,** *12* (9), 8970-8976.

46. Li, S.; Lin, Y. C.; Zhao, W.; Wu, J.; Wang, Z.; Hu, Z.; Shen, Y.; Tang, D. M.; Wang, J.; Zhang, Q.; Zhu, H.; Chu, L.; Zhao, W.; Liu, C.; Sun, Z.; Taniguchi, T.; Osada, M.; Chen, W.; Xu, Q. H.; Wee, A. T. S.; Suenaga, K.; Ding, F.; Eda, G., Vapour-liquid-solid growth of monolayer MoS$_2$ nanoribbons. *Nat Mater* **2018,** *17* (6), 535-542.

47. Mounet, N.; Gibertini, M.; Schwaller, P.; Campi, D.; Merkys, A.; Marrazzo, A.; Sohier, T.; Castelli, I. E.; Cepellotti, A.; Pizzi, G.; Marzari, N., Two-dimensional materials from high-throughput computational exfoliation of experimentally known compounds. *Nat Nanotechnol* **2018,** *13* (3), 246-252.